\documentclass[prl,aps,amssymb,showpacs,twocolumn]{revtex4}

\usepackage{dcolumn}
\usepackage{graphicx}
\usepackage{bm}

\usepackage{amsmath}
\usepackage{amssymb}
\usepackage{amsthm}
\usepackage{amsfonts}
\usepackage{enumerate}
\usepackage{latexsym}

\input{epsf}

\begin{document}
\title{ Pairing strengths for a two orbital model of the Fe-pnictides}
\author{Xiao-Liang Qi$^1$, S. Raghu$^1$, Chao-Xing Liu$^{2,1}$, D. J. Scalapino$^3$ and Shou-Cheng Zhang$^1$}
\affiliation{$^1$Department of Physics, McCullough Building,
Stanford University, Stanford, CA 94305-4045}
 \affiliation{$^2$Center for Advanced
Study, Tsinghua University, Beijing, 100084, R. P. China}
\affiliation{$^3$Department of Physics, University of California,
Santa Barbara, CA 93106-9530}
\date{\today}
\begin{abstract}
Using an RPA approximation, we have calculated the strengths of
the singlet and triplet pairing interactions which arise from the
exchange of spin and orbital fluctuations for a 2-orbital model of
the Fe-pnictide superconductors. When the system is doped with F,
the electron pockets become dominant and we find that the
strongest pairing occurs in the singlet d-wave pairing and the
triplet p-wave pairing channels, which compete closely. The
pairing structure in the singlet d-wave channel corresponds to a
superposition of near neighbor intra-orbital singlets with a minus
sign phase difference between the $d_{xz}$ and $d_{yz}$ pairs. The
leading pairing configuration in the triplet channel also involves
a nearest neighbor intra-orbital pairing.  We find that the
strengths of both the singlet and triplet pairing grow, with the
singlet pairing growing faster, as the onsite Coulomb interaction
approaches the value where the $S=1$ particle-hole susceptibility
diverges.

\end{abstract}


\maketitle

Recently, a new class of superconductors involving a family of
Fe-based oxypnictides has been discovered\cite{Kamihara2008,
ren2008, Chen2008, Chen2008A, Chen2008B, wen2008, ren2008A,
ren2008B, Lorenz2008}.  With $T_c$ as high as $55K$\cite{ren2008A},
the mechanism of superconductivity is likely to be electronic in
origin and consequently, these materials have generated tremendous
excitement.  Moreover, experimental results including specific
heat\cite{Mu2008,ding2008}, point-contact spectroscopy\cite{Shan2008},
high-field resistivity\cite{Hunte2008, Zhu2008} and NMR \cite{ahilan2008} measurements suggest
the existence of unconventional superconductivity in these
materials. Furthermore, transport\cite{dong2008} and neutron
scattering\cite{delacruz2008} measurements in LaOFeAs have shown the
evidence of spin-density-wave (SDW) magnetic order below $T=137K$.
An experimental determination of the orbital and spin state of the
Cooper pairs, however, has not yet been made.

Band structure calculations show that the Fermi surface of F doped
LaOFeAs consists of two nearly concentric hole cylinders
surrounding the $\Gamma$ point and two elliptically distorted
electron cylinders around the M point of the 2Fe/cell Brillouin
zone. Electronic transitions involving states on one or between
two of these Fermi surface sheets lead to q-dependent structure in
the spin and orbital susceptibilities. For small doping, the
electron and the hole fermi surfaces are of comparable sizes, and
their nesting can give rise to the observed SDW order in the
undoped material \cite{dong2008, delacruz2008}. Upon further
doping, the two hole pockets shrink, the electron fermi surfaces
become dominant \cite{zhang2008}, and the system exhibits
superconductivity. Ref.  \cite{Xu2008} suggests that a triplet
p-wave pairing state is obtained on the electron Fermi surfaces
due to the ferromagnetic spin fluctuations. Other related
possibilities have also been discussed in the literature,
including inter-orbital on-site triplet pairing \cite{dai2008,
lee2008}, and a s-wave pairing state which changes sign from the
electron to the hole pockets \cite{mazin2008}.

Recently, we have introduced a tight-binding model
\cite{raghu2008} with ``$d_{xz}$" and `` $d_{yz}$" orbitals on a
two-dimensional square lattice of ``Fe" sites. This simple
tight-binding model correctly reproduces the topology of both the
electron and the hole fermi surfaces. It also reproduces the van
Hove singularities obtained in bandstructure calculations. For low
doping, when the electron and hole pockets are comparable, RPA
calculations show enhanced SDW fluctuations at the wave vectors
$(\pi,0)$ and $(0,\pi)$, defined in the convention of one Fe atom
per unit cell \cite{raghu2008}. In this work we investigate the
nature of the pairing state when this model is further doped. With
on-site inter-orbital and intra-orbital Coulomb interaction terms,
we use the RPA approximation to study the effective pair
interaction vertex induced by the spin and orbital fluctuations.
We find that when the doping is increased and the electron pockets
become larger, the leading pairing instability occurs in the
singlet d-wave and the triplet p=wave channels. The pairing
strength for both channels increases as the system approaches an
instability in the $S=1$ particle-hole channel, with the singlet
d-wave channel growing faster than the triplet p-wave channel.

\textit{Model Hamiltonian - }Our tight-binding model Hamiltonian
describes a square two-dimensional ``Fe" lattice with two orbitals
per site
\begin{equation}
H_0=\sum_{k\sigma}\psi^+_{k\sigma}\left[\left(\varepsilon_+(k)-\mu\right)1+\varepsilon_-(k)
\tau_3+\varepsilon_{xy}(k)\tau_1\right]\psi_{k\sigma} \label{Htb}
\end{equation}
Here $\sigma$ is the spin index, $\tau_i$ are Pauli matrices and
$\psi^{\dagger}_{k\sigma}=
[d^{\dagger}_{x\sigma}(k),d^{\dagger}_{y\sigma}(k)]$ is a
two-component field, which describes the two degenerate ``$d_{xz}$''
and ``$d_{yz}$'' orbitals.  The matrix elements of $H_0$,
$\epsilon_+({\bf k})=-(t_1+t_2)(\cos k_x+\cos k_y)-4t_3\cos k_x\cos
k_y$, $\epsilon_-({\bf k})=-(t_1-t_2)(\cos k_x-\cos k_y)$ and
$\epsilon_{xy}({\bf k})=-4t_4\sin k_x\sin k_y$ are parametrized by
four hopping paramters $t_i, i=1,\cdots,4$.

\begin{figure}[ht]
\includegraphics [width=4.5cm,clip,angle=0]{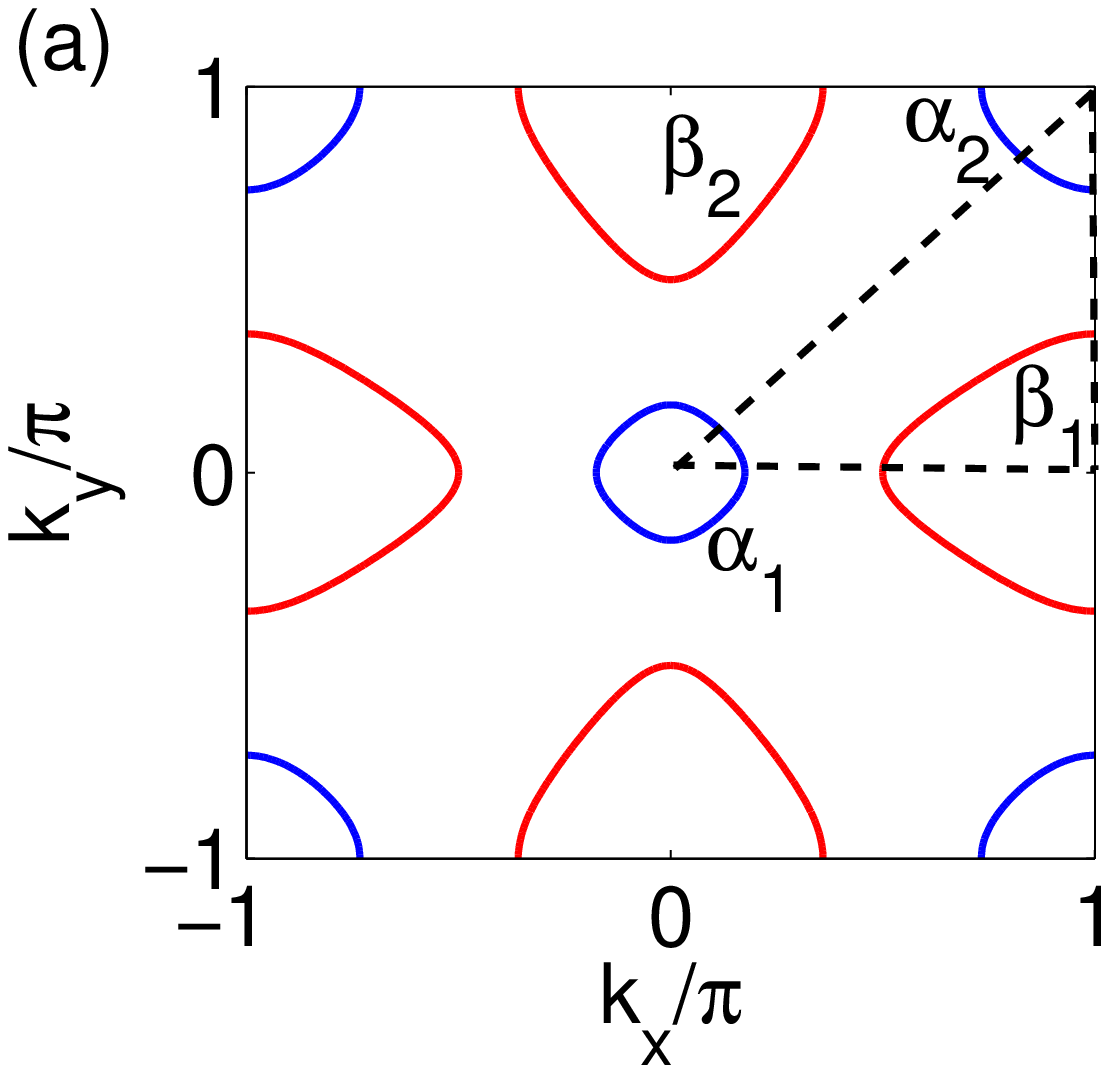}\includegraphics [width=4.5cm,clip,angle=0]{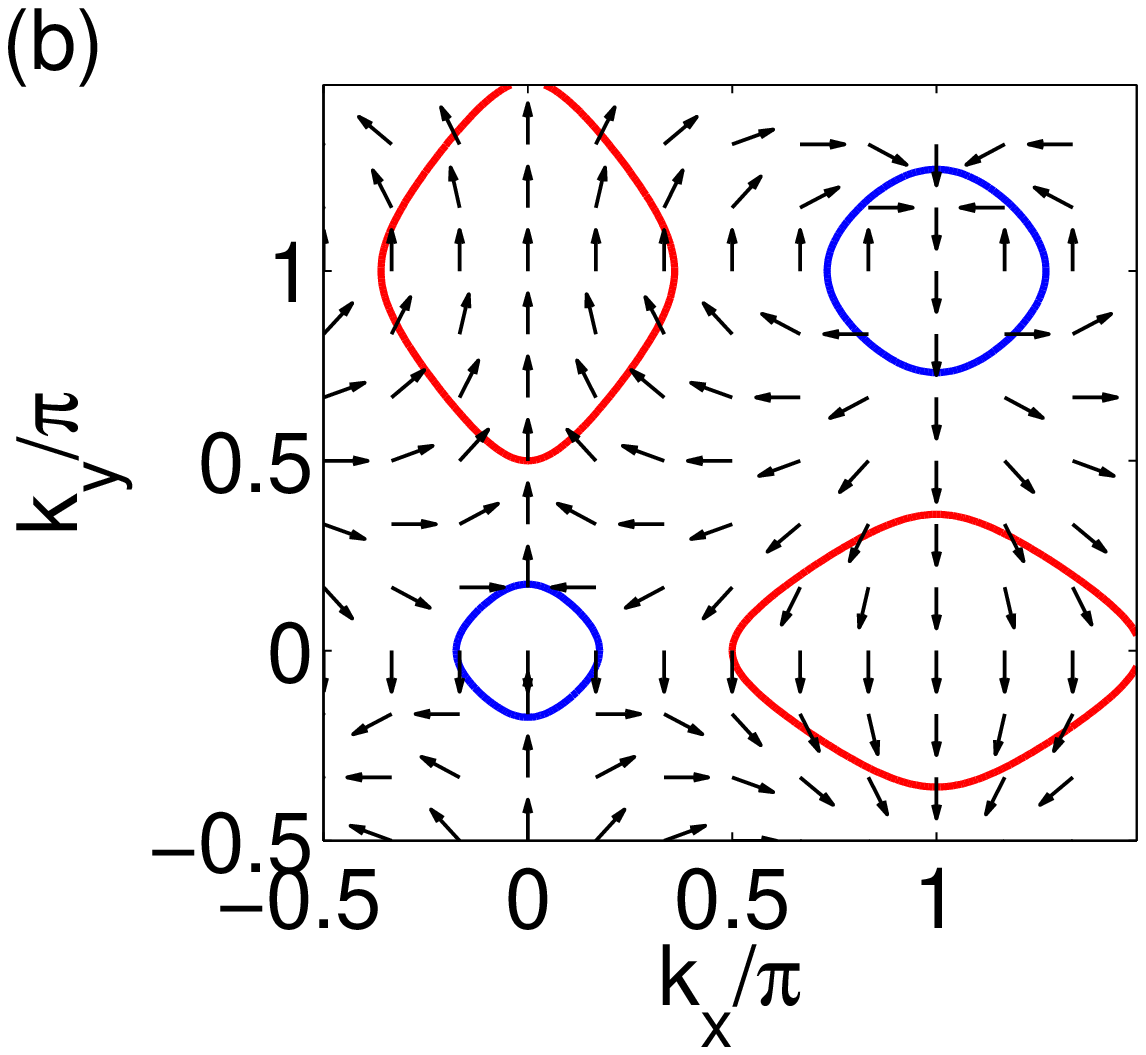}
\caption{(a) The Fermi surface of the 2-orbital model on the large
1Fe/cell BZ. Here, the $\alpha_{1,2}$ surfaces are hole Fermi pockets
given by $E_-(\bm k_f) = 0$ and the $\beta_{1,2}$ surfaces are electron Fermi
pockets given by $E_+(\bm k_f) = 0$. In this paper we have set
$t_1=-1,t_2 = 1.3, t_3=t_4=-0.85$ and $\mu = 2$. (b) Wave function
distribution in the Brillouin zone. The arrows show the direction of
the vector $(\epsilon_-({\bf k}),\epsilon_{xy}({\bf k}))$. When an
arrow is pointing up (down) at some ${\bf k}$ point, the eigenstate
of upper band $E_+({\bf k})$ consists of pure $xz$ ($yz$) orbitals.
The Brillouin zone is shifted by $(\pi/2,\pi/2)$ for convenience.}
\label{fs}
\end{figure}

This free fermion Hamiltonian is diagonalized by introducing a
canonical transformation to the band operators $\gamma_{\nu
\sigma,{\bf k}}$:
\begin{eqnarray}
\psi_{s\sigma,{\bf k}}=\sum_{\nu=\pm} a^s_{\nu,{\bf
k}}\gamma_{\nu\sigma,{\bf k}}\label{Bogolubov}
\end{eqnarray}
with
\begin{eqnarray}
a_{\nu,{\bf k}}^s&=&\left\langle s|\nu,{\bf k}\right\rangle \nonumber \\
a^x_{+ ,\bm k} &=& a^y_{-,\bm k}={\rm sgn}(\epsilon_{xy}(\bm
k))\sqrt{\frac{1}{2} + \frac{\epsilon_-(\bm
k)}{2\sqrt{\epsilon^2_-(\bm k) +
\epsilon^2_{xy}(\bm k)}}} \nonumber \\
a^y_{+, \bm k} &=& -a^x_{-,\bm k}=\sqrt{\frac{1}{2} -
\frac{\epsilon_-(\bm k)}{2\sqrt{\epsilon^2_-(\bm k) +
\epsilon^2_{xy}(\bm k)}}} \label{parameter}
\end{eqnarray}
the wave-function of the $\nu$ band with $\nu=\pm 1$, and
$\gamma_{\nu\sigma,{\bf k}}$ the annihilation operator of an
electron with spin $\sigma$ and wave-vector $\bm k$ in the $\nu$
band.
With the inclusion of a chemical potential $\mu$, the band part of
the Hamiltonian becomes
\begin{equation}
H_0 = \sum_{\bm k \sigma \nu} \left( E_{\nu} (\bm k) - \mu \right)
\gamma^{\dagger}_{\nu \sigma,{\bf k}}\gamma_{\nu \sigma,{\bf k}},
\qquad \nu = \pm
\end{equation}
with $E_{\pm} (\bm k) = \epsilon_{+}(\bm k) \pm
\sqrt{\epsilon^2_-(\bm k) + \epsilon^2_{xy}(\bm k)}$. The tight
binding parameters $t_i$ can be adjusted to fit the Fermi surface
obtained from LDA band structure calculations
\cite{Singh2008,mazin2008,Xu2008}.  In this work, we will take the
parameters $t_1=-1,t_2 = 1.3, t_3=t_4=-0.85$ and measure energy in
units of $\vert t_1 \vert$.
With a chemical potential $\mu=1.45$, one has a filling of 2
electrons per site, a Fermi surface similar to bandstructure
calculation of lightly doped LaOFeAs and a peak in the bare spin
susceptibility for $q=(\pi,0)$ and $(0,\pi)$. Here we will take
$\mu=2.0$ which corresponds to having 2.32 electron persite and
gives the Fermi surface shown in Fig. \ref{fs} (a). There are
four Fermi pockets in the Brillouin zone: $\alpha_1$
around $(0,0)$ and $\alpha_2$ around $(\pi,\pi)$ are hole
pockets associated with $E_-(\bm k_f) = 0$, while $\beta_1$
around $(\pi,0)$ and $\beta_2$ around $(0,\pi)$ are electron
pockets given by $E_+(\bm k_f) = 0$. For this minimal model, we will
include only onsite intra and inter orbital Coulomb interactions,
which will both be set equal to U and we will neglect the Hunds rule
coupling. In this case, up to a shift of the chemical potential, the
interaction can be written as
\begin{eqnarray}
    \hat{H}_{int}=\frac{U}{2}\sum_i\left(\sum_{\sigma}\psi^{\dag}_{\sigma}
    (i)\bold{1}\psi_{\sigma}(i)\right)^2
    \label{Hint}
\end{eqnarray}

An important feature of this two band model is the nontrivial $C_4$
rotation symmetry of the two orbitals. Under a $90^\circ$ degree
rotation, the two orbitals transform as $|xz\rangle\rightarrow
|yz\rangle$ and $|yz\rangle\rightarrow -|xz\rangle$.
Correspondingly, in the Hamiltonian, $\epsilon_+({\bf k})$ has
$s$-wave symmetry and $\epsilon_-({\bf k})$ and $\epsilon_{xy}({\bf
k})$ have $d$-wave symmetry, which together perserve the point
group symmetry of the Hamiltonian.  Consequently, the
wave functions $a_{\nu,{\bf k}}^s$ of the energy eigenstates also
have nontrivial structure in the Brillouin zone, which can be
determined by the direction of the vector ${\bf n}({\bf
k})=(\epsilon_-({\bf k}),\epsilon_{xy}({\bf k}))$. If we consider
the orbital degree of freedom as a pseudo-spin, the electrons in the
lower band always have a ``pseudo-spin" anti-parallel to ${\bf
n}({\bf k})$. For the parameters we are using, the distribution of
the unit vector ${\bf \hat{n}}({\bf k})={\bf n}({\bf k})/|{\bf
n}({\bf k})|$ is shown in Fig. \ref{fs} (b). For example, at
the wavevector ${\bf k}=(\pi/2,0)$ we have $\epsilon_-({\bf k})<0$
and $\epsilon_{xy}({\bf k})=0$, which means the upper band is formed
from $xz$ orbitals and the lower band from $yz$ orbitals. From Fig.
\ref{fs} (b) we can see that the electron pocket $\beta_1$
($\beta_2$) is formed mainly from $xz$ ($yz$) orbitals, while the
hole pockets are formed from ``$d$-wave" superposition of the two
orbitals. This point will be important for understanding the pairing
symmetry.
Since this nontrivial structure of the wave function originates from
the symmetry of the two orbitals $d_{xz}$ and $d_{yz}$, we expect it
to be qualitatively correct even beyond the present two orbital
model.

In the following we first discuss the particle-hole susceptibility
and calculate it using an RPA approximation. Then using the pairing
interaction associated with the exchange of these particle-hole
excitations, we examine the strength of the pairing in the singlet and
triplet channels.

\textit{One loop and RPA susceptibilities - } Because of the
two-orbitals, the generic form of the susceptibility depends on four
orbital indices $p,q,s,t$ equal to $1$ or $2$ for $d_{xz}$ and
$d_{yz}$, as well as spin indices:
\begin{eqnarray}
\chi_{s\alpha,t\beta}^{p\gamma,q\delta}({\bf q},i\Omega)&=&\int
\frac{d^2{\bf k}}{(2\pi)^2}\int_0^\beta d\tau e^{i\Omega
\tau}\left\langle T_\tau \psi_{t\beta,{\bf
k-q}}^\dagger(\tau)\right.\nonumber\\
& &\cdot\psi_{s\alpha,{\bf k}}(\tau)\left.\psi_{p\gamma,{\bf
k'+q}}^\dagger(0)\psi_{q\delta,{\bf k'}}(0)\right\rangle
\end{eqnarray}
Due to the $SU(2)$ spin rotation symmetry, the susceptibility
function has the following form:
\begin{eqnarray}
\chi_{s\alpha,t\beta}^{p\gamma,q\delta}({\bf
q},i\Omega)&=&\frac16{\chi_1}_{st}^{pq}\vec{\sigma}_{\beta\alpha}\cdot\vec{\sigma}_{\gamma\delta}+
\frac12{\chi_0}_{st}^{pq} \delta_{\beta\alpha}\delta_{\gamma\delta}
\end{eqnarray}
where $\chi_1$ and $\chi_0$ correspond to the correlation functions
of the triplet fields (such as spin) and the singlet fields (such as
charge density), respectively. All the physical susceptibilities are
determined by some components of ${\chi_{0,1}}_{st}^{pq}$. For
example, the total spin susceptibility is given by
$\chi_S=\frac12\sum_{s,p}{\chi_1}_{ss}^{pp}$. At the one-loop level,
we have ${\chi_0}_{st}^{pq}({\bf q},i\Omega)={\chi_1}_{st}^{pq}({\bf
q},i\Omega)$, which we denote by $\chi_{st}^{pq}({\bf q},i\Omega)$.
For a given $({\bf q},i\Omega)$, ${\chi_{0}}_{st}^{pq}$ and
${\chi_1}_{st}^{pq}$ are $4\times 4$ matrices, and the RPA
susceptibility is obtained from the matrix equation
\begin{eqnarray}
\chi_{0(1)}^{\rm RPA}({\bf q},i\Omega)=\chi({\bf
q},i\Omega)\left(\mathbb{I}-\gamma_{0(1)}\chi({\bf
q},i\Omega)\right)^{-1}
\end{eqnarray}
with
\begin{eqnarray}
\gamma_1=U\mathbb{I}_{4\times
4},~~\gamma_0=\left(\begin{array}{cccc}-U&&&\\&U&&\\&&U&\\&&&-U\end{array}\right),
\end{eqnarray}
in the basis $(st)=(11,21,12,22)$. The one-loop susceptibility
$\chi_{st}^{pq}({\bf q},i\Omega)$ is given by
\begin{eqnarray}
\chi_{st}^{pq}({\bf q};i\Omega)&=&-\int\frac{d^2{\bf
k}}{(2\pi)^2}\frac{a_{\nu,{\bf k+q}}^{t*}a_{\nu',{\bf
k}}^sa_{\nu',{\bf k}}^{p*}a_{\nu,{\bf k+q}}^q}
{i\Omega+E_{\nu,{\bf k+q}}-E_{\nu',{\bf k}}}\nonumber\\
& &\cdot\left({n_F({E_{\nu,{\bf k+q}}})-n_F({E_{\nu',{\bf
k}}})}\right)\label{bubble}
\end{eqnarray}
with $a_{\nu,{\bf k}}^s$ defined by Eq. (\ref{parameter}).

\begin{figure}[ht]
\includegraphics
[width=4.3cm,clip,angle=0]{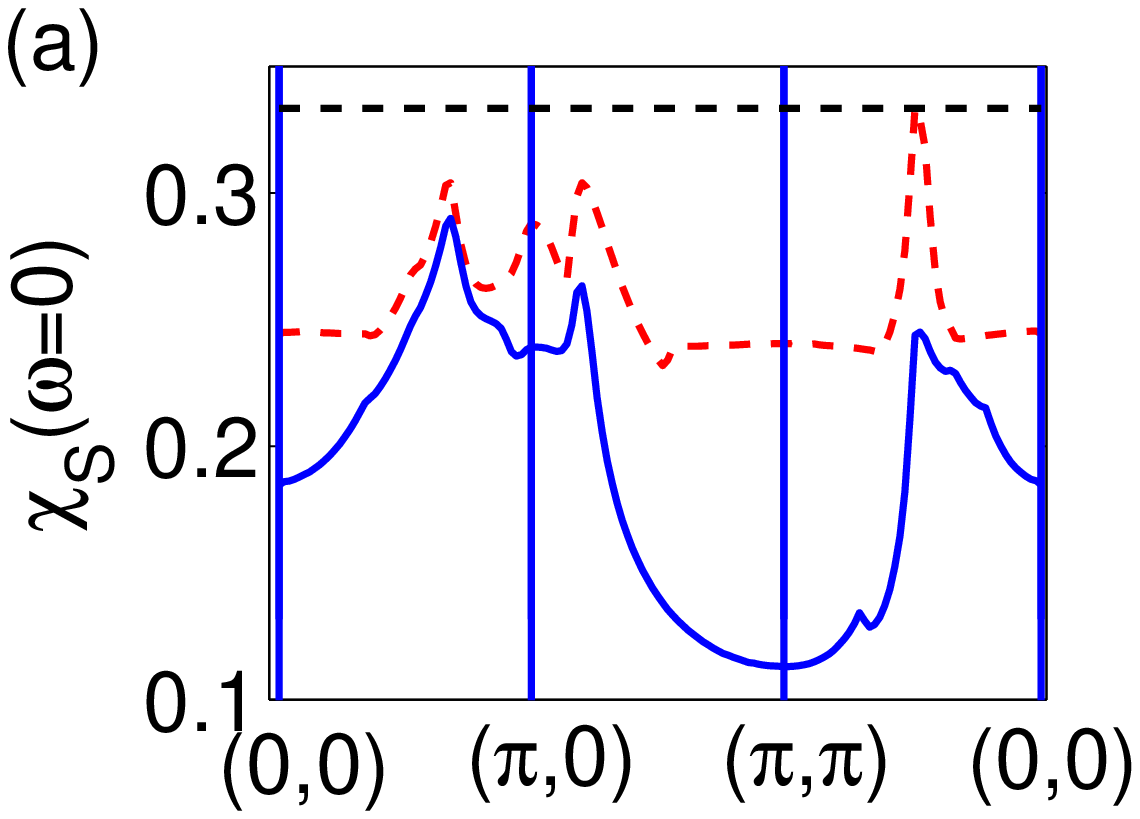}\includegraphics[width=4.3cm,clip,angle=0]{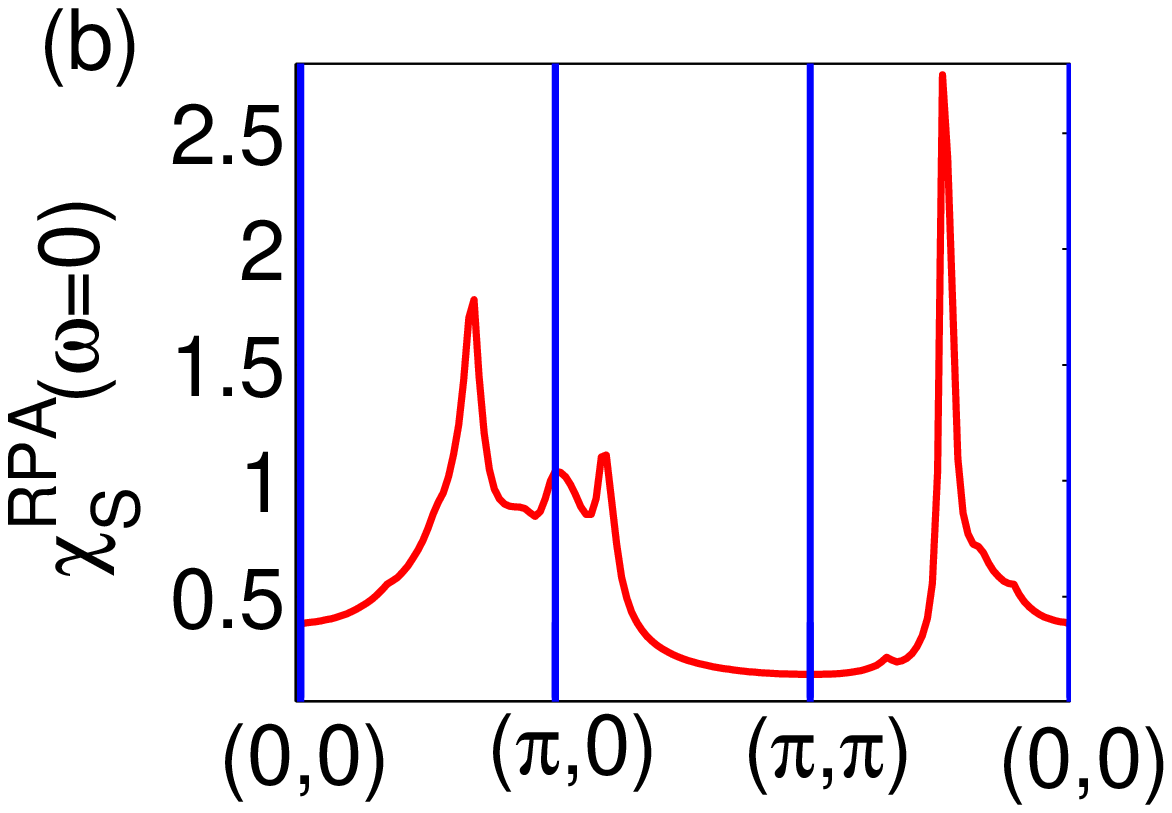}
\caption{(a) The static one loop spin susceptibility (solid line)
and the largest eigenvalue of one loop susceptibility matrix
$\chi_{st}^{pq}$ along the $(0,0)\rightarrow (\pi,0)\rightarrow
(\pi,\pi)\rightarrow (0,0)$ contour in the Brillouin zone. The
horizontal dashed line shows the value of $1/U_c=1/3$ which
indicates the critical $U_c=3$. (b) The RPA spin susceptibility for
$U=2.8$.} \label{chi0}
\end{figure}

In Fig. \ref{chi0} (a), the one loop spin susceptibility
$\chi_S({\bf q},\omega=0)$ versus momentum ${\bf q}$ for $\mu=2.0$
is shown as the solid curve. The dashed curve shows the maximal
eigenvalue of the one-loop susceptibility matrix $\chi_{st}^{pq}$
along the same contour. From this, we see that there is a critical
value $U_c\simeq 3$, at which the $S=1$ generalized RPA
susceptibility diverges at an incommensurate wave vector near
${\bf q}\simeq(\pi/2,\pi/2)$. This divergence occurs in the
spin-one part of the particle-hole channel and reflects a
superposition of particle-hole spin-one fluctuations involving
both orbitals. The RPA spin susceptibility for $U=2.8$ is also
shown in Fig. \ref{chi0} (b), which, as expected, shows the
strongest enhancement near ${\bf q}=(\pi/2,\pi/2)$.

\textit{Superconductivity - } Within an RPA
approximation, the singlet and triplet pairing vertices are given
by\cite{Takimoto2004}
\begin{eqnarray}
{\Gamma_0}_{st}^{pq}({\bf
k,k'},i\Omega)&=&-\frac12\left(U_0-3U_1\right)_{ps}^{tq}({\bf
k-k'},i\Omega)\nonumber\\
{\Gamma_1}_{st}^{pq}({\bf
k,k'},i\Omega)&=&-\frac12\left(U_0+U_1\right)_{ps}^{tq}({\bf
k-k'},i\Omega)\label{Gamma01}
\end{eqnarray}
Here ${U_{0}}_{ps}^{tq}=\left[\frac12\gamma_0+\gamma_0\chi_{0}^{\rm
RPA} \gamma_0\right]_{ps}^{tq}$ and
${U_{1}}_{ps}^{tq}=\left[\frac12\gamma_1+ \gamma_1\chi_{1}^{\rm
RPA}\gamma_1\right]_{ps}^{tq}$ describe the effective interaction
mediated by orbital and spin fluctuation respectively. It should be
noticed that the order of orbital indices is different for $U_{0,1}$
and $\Gamma_{0,1}$.

Just as for the traditional phonon case, retardation is important
and what enters in characterizing the strength of the pairing
interaction is
\begin{equation}
\int_{0}^{\infty} \frac{d \omega}{\pi}  \frac{{\rm Im} \left[
{\Gamma_{0(1)}}^{pq}_{st}\left( \bm k, \bm k',\omega \right)
\right]}{\omega} = {\rm Re}\left[{\Gamma_{0(1)}}_{st}^{pq}\left(\bm
k, \bm k', \omega=0 \right)\right]
\end{equation}
in which a Wick rotation $i\Omega\rightarrow \omega+i\delta$ has
been performed on $\Gamma_{0,1}({\bf k,k'},i\Omega)$. The
interaction induces scattering of two Cooper pairs around the Fermi
surfaces. For later convenience, we define $C_i,i=1,...,4$ as the
four pieces of Fermi pockets $\alpha_1,\alpha_2,\beta_1,\beta_2$,
and $\gamma_{i\sigma {\bf k}}$ the annihilation operator of the
electron around the $i^{\rm th}$ Fermi surface pocket with
wavevector ${\bf k}\in C_i$. Thus $\gamma_{i\sigma {\bf k}}$ is
equal to $\gamma_{\nu_i,\sigma {\bf k}}$ defined in Eq.
(\ref{Bogolubov}) with $\nu_i=+1(-1)$ when $C_i$ is an electron
(hole) pocket. The Cooper pair defined here can be either a singlet
or a triplet. Here and below we omit the spin indices since the spin
state is determined by the parity of the gap when ${\bf k}$ goes to
${\bf -k}$. The scattering of a Cooper pair from $({\bf k},-{\bf
k})$ on the ith Fermi surface to $({\bf k}',-{\bf k}')$ on the jth
Fermi surface is determined by the projection of the interaction
vertex ${\Gamma_{0,1}}_{st}^{pq}({\bf k,k'})$ to the energy
eigenstates:
\begin{eqnarray}
{\Gamma_{0,1}}_{ij}({\bf k,k'})=\sum_{s,t,p,q} a^{t*}_{\nu_i,-\bf
k}a^{s*}_{\nu_i,\bf k} {\Gamma_{0,1}}^{pq}_{st} (\bm k, \bm k')
a^p_{\nu_j,{\bf k}}a^q_{\nu_j,-{\bf k}}\label{Gammaij}
\end{eqnarray}
with ${\bf k}\in C_i,~{\bf k'}\in C_j$.

For a pairing configuration mediated by $\Delta({\bf k})=g({\bf
k})\gamma_{i,-{\bf k}}\gamma_{i,{\bf k}},~{\bf k}\in C_i$, a
dimensionless {\em coupling strength functional} is defined
as\cite{Scalapino1986}
\begin{eqnarray}
\lambda[g({\bf k})] = -\frac{\sum_{i,j}\oint_{C_i} \frac{d{\bf
k}_\parallel}{v({\bf k})}\oint_{C_{j}}\frac{d{\bf
k'}_\parallel}{v({\bf k'})} g(\bm k) \Gamma^{[g]}_{ij}(\bm k, \bm
k') g(\bm k') } {(2\pi)^2\sum_{i}\oint_{C_i} \frac{d{\bf
k}_\parallel}{v({\bf k})} g^2(\bm k)}\label{lambda}
\end{eqnarray}
in which $v({\bf k})=|\nabla_{\bf k}E_{\nu(i)}({\bf k})|$ for ${\bf
k}\in C_i$ is the fermi velocity, and $\oint_{C_i}\frac{d{\bf
k}_\parallel}{v({\bf k})}$ is a loop integral around the $C_i$ fermi
surfaces. ${\Gamma^{[g]}_{ij}}({\bf k,k'})={\Gamma_{0(1)}}_{ij}({\bf
k,k'})$ when $g({\bf k})$ has even (odd) parity, respectively. For a
given ${\Gamma_{0,1}}_{ij}({\bf k,k'})$, the optimum pairing
configuration and corresponding $\lambda$ can be determined by
solving an eigenvalue problem
\begin{eqnarray}
-\sum_{j}\oint_{C_j}\frac {d{\bf k'}_\parallel}{(2\pi)^2 v({\bf
k'})}\Gamma^{[g]}_{ij}({\bf k,k'})g({\bf k'})=\lambda g({\bf
k}),\label{Eiglambda}
\end{eqnarray}
which is obtained from the stationary condition
$\delta\lambda[g({\bf k})]/\delta g({\bf k})=0$.

\begin{figure}[ht]
\includegraphics
[width=8.8cm,clip,angle=0]{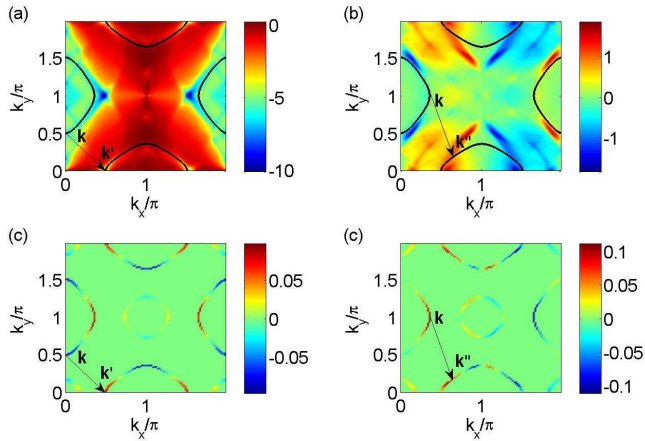} \caption{(a) The effective
interaction ${\Gamma_{0}}_{++}({\bf k,k'})$ describing the
scattering of singlet Cooper pairs on and between electron pockets.
One wavevector ${\bf k}$ is fixed and the color shows the value of
${\Gamma_{0}}_{++}({\bf k,k'})$ as a function of ${\bf k'}$. (b) The
same plot as (a) for the triplet channel $\Gamma_1({\bf k,k'})$. (c)
and (d) shows the optimum singlet and triplet pairing
configurations. The arrow in each figure indicates a typical inter
Fermi surface scattering process.} \label{figGamma}
\end{figure}
The interaction ${\Gamma_{0,1}}_{ij}({\bf k,k'})$ contains various
intra Fermi surface and inter Fermi surface scattering processes.
However, for $\mu=2.0$ we find that scattering on and between the
$\beta_1$ and $\beta_2$ electron pockets is dominant. From the
definition (\ref{Gammaij}) we see that ${\Gamma_{0,1}}_{ij}$ only
depends on the band label $\nu_i,\nu_j$. Thus the scattering on and
between the two electron pockets are determined by ${\Gamma_{0,1}}_{++}({\bf k,k'})$
Eq. (\ref{Gammaij}), since $\nu_i=\nu_j=+1$ (which we have shortened to $+$).
The distribution of the effective singlet $\Gamma_{0++}(k,k')$ and triplet
$\Gamma_{1,++}(k,k')$ interactions for $U=2.8$ are
shown in Fig. \ref{figGamma} (a) and (b) respectively for a fixed ${\bf k}\in \beta_2$.
From the figure it can be seen that the interaction for singlet
pairing is repulsive, which favors a pairing configuration with
a node. The interaction in the triplet channel has a smaller
magnitude than the singlet, but is still sizable and can support a
p-wave triplet state. The relative sign of the order parameter on
the two fermi pockets is determined by the inter Fermi surface
scattering. For example, ${\Gamma_0}_{++}({\bf k,k'})$ is repulsive
for the choice of ${\bf k}$ and ${\bf k'}$ shown by the arrow in
Fig. \ref{figGamma} (a) and (c), so that the pairing amplitude has
opposite signs at ${\bf k}$ and ${\bf k'}$. For a similar reason the
pairing amplitude in Fig. \ref{figGamma} (b) and (d) has the same
sign at ${\bf k}$ and ${\bf k''}$.

By solving Eq. (\ref{Eiglambda}), we obtain the optimum singlet and
triplet pairing configurations shown in Fig. \ref{figGamma} (c)
and (d) respectively. We find that $\lambda_0=0.46$ for singlet pairing and
$\lambda_1=0.20$ for triplet pairing. To see the contributions of
intra Fermi surface and inter Fermi surface scattering processes, we
calculate these two terms separately by defining $\lambda_{ij}$
as the term in Eq. (\ref{lambda}) that involves the scattering
from the $i$ to the $j$ Fermi surface. The total $\lambda$ is given by
$\lambda=\sum_{i,j}\lambda_{ij}$. For $i,j=\beta_1,\beta_2$,
${\lambda_{0,1}}_{ij}$ are $2\times 2$ matrices, which for $U=V=2.8$
are
\begin{eqnarray}
\lambda_0=\left(\begin{array}{cc}     0.15
          &  0.083\\0.083  &
0.15\end{array}\right),~ \lambda_1=\left(\begin{array}{cc} -0.026
        &
0.10\\0.10 & 0.025\end{array}\right)
\end{eqnarray}
Here one see that the inter Fermi surface scattering makes an
important contribution to the pairing strength $\lambda$. It should
be noticed that the singlet pairing configuration has
the same diagonal term $\lambda_{ii}$ for
$i=\beta_1$ and $\beta_2$ while the p-wave diagonal terms
$\lambda_{ii}$ can have different values.

\begin{figure}[ht]
\includegraphics
[width=4.5cm,clip,angle=0]{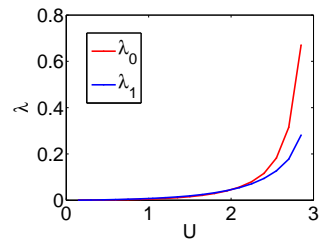} \caption{ The singlet
$\lambda_0$ and triplet $\lambda_1$ pairing strength as a function of
$U$.}
\label{lambdavsU}
\end{figure}

We have also studied the dependence of the pairing strength upon the
interaction $U$. As shown in Fig. \ref{lambdavsU}, the strength of
both the singlet d-wave and triplet p-wave pairing channels is
increased as U increases. Within the RPA approximation, the coupling
strength formally diverge as U approaches the critical point $U_c\approx 3.0$
associated with the onset of order in the spin one particle-hole
channel. This is similar to the behavior found for the
$d_{x^2-y^2}$ coupling in the two-dimensional Hubbard model when an
RPA approximation is used to treat the pairing due to the exchange
coupling of spin-flucturations. Just as for that case, one needs to
go beyond the RPA to determine the actual behavior of the model.

In order to gain further insight into the nature of the pairing,
it is useful to determine the real space pairing
structure which corresopnds to $\Delta({\bf k})$ shown in
Fig. \ref{figGamma} (c) and (d). This can be obtained from
$\Delta^{\dag}=\sum_k g(k)\gamma^{\dag}_{+\uparrow}(k)\gamma^{\dag}
_{+\downarrow}(-k)$ by transforming the band opertor to orbital operators,
Eq (\ref{parameter}), and Fourier transforming to the lattice coordinates.
As discussed earlier, the states associated with the
two electron pockets $\beta_1$ and $\beta_2$ are formed primarily
from $xz$ and $yz$ orbitals, respectively, as shown in Fig.
\ref{fs} (b).
With this in mind, we find that the singlet
pairing in Fig 5(c) corresponds to a superposition of singlets
formed from electrons in near-neighbor $d_{xz}$ orbitals mimus a
similar super position involving the $d_{yz}$ orbitals.
\begin{eqnarray}
    &&\Delta^{\dag}_{d}=\frac{1}{N}\sum_{r,\hat{\delta}=\hat{x},\hat{y}}
    (\psi^{\dag}_{x\uparrow}(r)\psi^{\dag}_{x\downarrow}(r+\hat{\delta})
    -\psi^{\dag}_{x\downarrow}(r)\psi^{\dag}_{x\uparrow}(r+\hat{\delta}))
    \nonumber\\
    &&-(\psi^{\dag}_{y\uparrow}(r)\psi^{\dag}_{y\downarrow}(r+\hat{\delta}
    )-\psi^{\dag}_{y\downarrow}(r)\psi^{\dag}_{y\uparrow}(r+\hat{\delta}))
    \label{}
\end{eqnarray}
Under a $90$ degree rotation
$\psi^{\dag}_{x\sigma}\rightarrow\psi^{\dag}_{y\sigma}$ and
$\psi^{\dag}_{y\sigma}\rightarrow -\psi^{\dag}_{x\sigma}$, so that
$\Delta^{\dag} _d$ changes sign, corresponding to a d-wave gap. For
the p-wave triplet shown in Fig 5(d), we find that
\begin{eqnarray}
    \Delta^{\dag}_{p}=\frac{1}{N}\sum_{r,\hat{\delta}=\hat{x},\hat{y}}(\psi^{\dag}
    _{x\uparrow}(r)\psi^{\dag}_{x\uparrow}(r+\hat{\delta})+\psi^{\dag}_{y\uparrow}
    (r)\psi^{\dag}_{y\uparrow}(r+\hat{\delta}))
    \label{}
\end{eqnarray}
Note that this triplet gap is associated with a near neighbor
intra-orbital pairing rather than the onsite inter-orbital triplet
pairing proposed in Ref. \onlinecite{lee2008}.

\textit{Conclusion - } We have studied the pairing interaction
associated with the exchange of particle-hole fluctuation for a
two-orbital $d_{xz}$-$d_{yz}$ Hubbard model. By adjusting the tight
binding parameters, one can obtain Fermi surface with hole and
electron pockets which are similar to those found in bandstructure
calculations for LaOFeAs. For a filling of two electrons per site,
the signs of the hole and electron pockets are similar and the RPA
spin susceptibility becomes singular as the on site intra and inter
Coulomb interaction U increase. This SDW singularity occurs at a
wave vector $q=(\pi,0)$ and $(0,\pi)$ associated with the nesting of
the hole and electron pockets. When this model is doped, the hole
pockets shrink and the electron pockets become dominant. In this
case, the SDW $q=(\pi,0)$ singularity in the spin susceptibility is
suppressed and there is a strong response in the spin one
particle-hole channel near $q\approx (\pi/2,\pi/2)$. The pairing
interaction associated with the exchange of these fluctuation leads
to an attractive interaction for both singlet d-wave and triplet
p-wave pairing, which compete closely. The singlet d-wave pairing
strength grows faster than the triplet p-wave pairing strength as
the interactions are increased and a magnetic instability is
approached. However, more refined numerical calculations beyond the
RPA approximation are needed to uniquely select among the two
competing pairing states.

\textit{Acknowledgement - } We acknowledge helpful discussions with
X. Dai, Z. Fang, S. Kivelson, T. Maier, R. Martin, I. Mazin, T.
Schulthess, D. Singh and H. Yao. This work is supported by the NSF
under grant numbers DMR-0342832, the US Department of Energy, Office
of Basic Energy Sciences under contract DE-AC03-76SF00515, the
center for nanophase material science, ORNL (DJS) and the Stanford
Institute for Theoretical Physics (SR, DJS).

\textit{Note added - } After completing this work, we learned that a
similar work has been done by Z.-J. Yao, J.-X. Li and Z. D. Wang,
which is posted in arXiv:0804.4116\cite{yao2008}.

\bibliography{LaOFeAs}

\end{document}